\documentclass[10pt]{article}

\usepackage{ametsoc2col}
\usepackage[colorlinks=true,
            linkcolor=blue,
            urlcolor=blue,
            citecolor=blue]{hyperref}

\makeatletter
\@ifpackageloaded{ametsoc}{
\DeclareGraphicsExtensions{.pdf} 
\newcommand{\mytable}{0.81 \linewidth}

}
{%
\DeclareGraphicsExtensions{.pdf} 
\graphicspath{{./ArXiv}}
\newcommand{\mytable}{0.65 \linewidth}

\usepackage{enumitem}

}
\makeatother

\usepackage{booktabs}
\usepackage{siunitx}
\usepackage{relsize}
\usepackage{pdflscape}
\usepackage[english]{babel}
\usepackage{rotating}
\usepackage{caption}

\newcommand{\myabstract}{Coastally associated rainfall is a common feature especially in tropical and subtropical regions. However, it has been difficult to quantify the contribution of coastal rainfall features to the overall local rainfall. We develop a novel technique to objectively identify precipitation associated with land-sea interaction and apply it to satellite based rainfall estimates. The Maritime Continent, the Bight of Panama, Madagascar and the Mediterranean are found to be regions where land-sea interactions play a crucial role in the formation of precipitation. In these regions $\approx$ 40\% to 60\% of the total rainfall can be related to coastline effects. Due to its importance for the climate system, the Maritime Continent is a particular region of interest with high overall amounts of rainfall and large fractions resulting from land-sea interactions throughout the year. To demonstrate the utility of our identification method we investigate the influence of several modes of variability, such as the Madden-Julian-Oscillation and the El Ni\~{n}o Southern Oscillation, on coastal rainfall behavior. The results suggest that during large scale suppressed convective conditions coastal effects tend to modulate the rainfall over the Maritime Continent leading to enhanced rainfall over land regions compared to the surrounding oceans. We propose that the novel objective dataset of coastally influenced precipitation can be used in a variety of ways, such as to inform cumulus parametrization or as an additional tool for evaluating the simulation of coastal precipitation within weather and climate models.}
\begin{document}
%
%
\title{\textbf{\large{Global detection and analysis of coastline associated rainfall using an objective pattern recognition technique}}}
%
%
\author{\textsc{Martin Bergemann}
				\thanks{\textit{Corresponding author address:} 
				Martin Bergemann, School of Earth, Atmosphere and Environment, Monash University, Melbourne, VIC 3800, Australia. 
				\newline{E-mail: martin.bergemann@monash.edu}}\quad\textsc{Christian Jakob}\\
\textit{\footnotesize{School of Earth, Atmosphere and Environment, Faculty of Science, Monash University, Melbourne, VIC 3800, Australia}} \\ \textit{\footnotesize{ARC Centre of Excellence for Climate System Science}}
\and 
\centerline{\textsc{Todd P. Lane}}\\
\centerline{\textit{\footnotesize{School of Earth Sciences, The University of Melbourne, Melbourne, VIC 3800, Australia}}} \\ \centerline{\textit{\footnotesize{ARC Centre of Excellence for Climate System Science}}}
}
%
\ifthenelse{\boolean{dc}}
{
\twocolumn[
\begin{@twocolumnfalse}
\amstitle

\begin{center}
\begin{minipage}{13.0cm}
\begin{abstract}
	\myabstract
	\newline
	\begin{center}
		\rule{38mm}{0.2mm}
	\end{center}
\end{abstract}
\end{minipage}
\end{center}
\end{@twocolumnfalse}
]
}
{
\amstitle
\begin{abstract}
\myabstract

\end{abstract}
\newpage
}
\section{Introduction}
\label{Sec_1}
Precipitation, one of the most important meteorological variables, is strongly affected by variations in solar forcing. As a result tropical rainfall variability is strongly dominated by the seasonal and the diurnal cycle. \cite{Yang2001} showed the importance of rainfall variance within diurnal and sub-diurnal frequencies for coastal tropical regions such as the Maritime Continent. In this area the diurnal rainfall variability is thought to be mostly generated by land-sea breeze circulations \citep{Mori2004}.
\begin{figure*}
\centering
\begin{minipage}{0.85\textwidth}
\includegraphics[width=\textwidth]{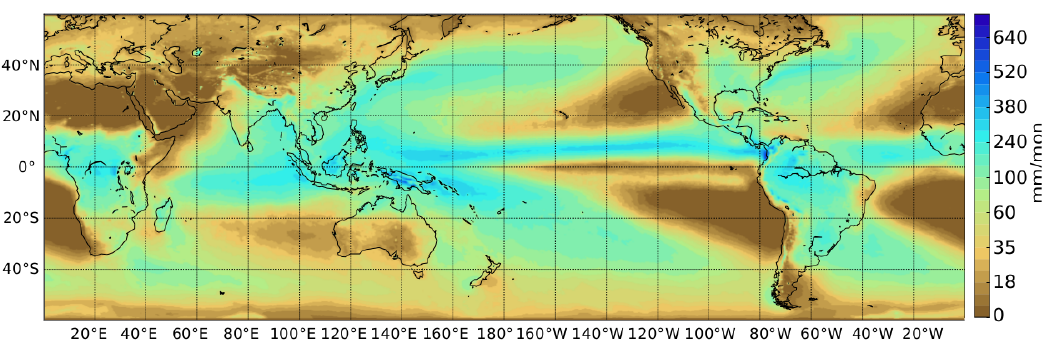}
\caption{Average monthly sum of total rainfall (1998 - 2013) for the region that is covered by the CMORPH satellite based rainfall estimates}
\label{totprecip_yearly}
\end{minipage}
\end{figure*}
Land-sea breeze systems are mainly forced by differential heating between land and the adjacent ocean but also affected by a variety of different factors such as coastline curvature, latitude, topography, atmospheric stability, land use and synoptic wind patterns \citep[e.g.][]{McPherson1970,Haurwitz1947,Pielke2002,Estoque1962,Mak1976,Mahrer1977}. \cite{Crosman2010} provide a  comprehensive review of the studies that have been conducted about the nature of land-sea breeze circulation systems. These circulation systems cause characteristic rainfall patterns in coastal regions. \cite{Mori2004} showed that the Maritime Continent rainfall between 2100 LT and 0900 LT is concentrated over the oceans peaking in the early morning. The 0900 LT to 2100 LT precipitation is mainly located over land with maxima occurring in the early evening. The rainfall patterns associated with land-sea interaction tend to propagate roughly 150 km on- and offshore \citep{Keenan2008}. Further propagation can occur through the interaction with other phenomena like mountain-valley breeze systems \citep{Qian2008} or gravity waves \citep{Mapes2003}.

Roughly {20 \%} of the world population live within the area that is affected by coastal precipitation. Additionally the mean population density is about three times higher  near coasts than on global average \citep{Small2003}. Coastal areas are also vulnerable to an increase of storm surges and heavy precipitation. Therefore a more accurate simulation of coastal rainfall in global climate models can potentially contribute directly to a better assessment of climate impacts on coastal areas. Precipitation on the other hand remains a challenging meteorological variable in general circulation models. Several studies have documented the issues in representing precipitation in climate models. For example \cite{Sun2006} and \cite{Dai2006} compared rainfall simulations of 18 coupled models with observations and found that while most models are able to capture the broad pattern of precipitation amount and year-to-year variability they fail to reproduce the diurnal cycle. Compared with observations  modeled rainfall is too weak and too frequent \citep{Stephens2010}. In coastal areas, the spatial pattern and timing of precipitation becomes worse \citep{Collier2004}. Especially over the Maritime Continent most climate models reveal a dry-wet rainfall bias, with too wet conditions over the ocean and too dry conditions over land or vice versa. It is likely that this is related to the complex structure of islands in combination with steep terrain. This combination leads to complex coastal convective systems causing rainfall patterns that are not easily captured by the relatively coarse-resolution global climate models. Although advances in convection parameterization and model resolution have been made to tackle the problem of rainfall timing and intensity in climate models the main issues in simulating the diurnal precipitation cycle remain unsolved \citep[e.g.][]{Mapes2003b,Slingo2004,Sato2009,Gianotti2011,Folkins2014}.

One of the key problems for an accurate description of coastal convection and precipitation is a fundamental lack of a global dataset of coastally induced precipitation. Although many studies have been conducted to describe convection, precipitation and land-sea-breeze circulation systems \citep[e.g.][]{
Frizzola1963,Tijm1999,Zhuo2013,Wapler2012}, the vast majority of them are both, local and phenomenological, because precipitation that is induced by land-sea interaction has to be separated from the background state of overall rainfall. So far, to our knowledge, no method exists that attempts to objectively identify rainfall directly associated with land-sea interactions. The aim of this study is to describe and evaluate a method that objectively finds coastal precipitation patterns that are related to land-sea interaction. Once the method is developed and assessed a global climatology of coastally induced rainfall and its diurnal cycle are presented. To give an example for the utility of the derived dataset, the role of the Madden-Julian-Oscillation (MJO) and the El Ni\~{n}o Southern Oscillation (ENSO) on coastal precipitation in the Maritime Continent region are then investigated.

\textcolor{blue}{Section} \ref{Sec_2} describes the rainfall data used in the study. \textcolor{blue}{Section} \ref{Sec_3}  discusses the broad features of the objective coastal rainfall detection technique, with a more technical description provided in in the \textcolor{blue}{appendix}. \textcolor{blue}{Section} \ref{Sec_4} to \ref{Sec_6} then describe its application to a global rainfall dataset to study the global climatology of coastal rainfall, its diurnal cycle and the influence of larger scale modes of variability on the coastal rainfall occurrence. This is followed by  a summary and conclusions in \textcolor{blue}{section} \ref{Sec_7}.

\section{Rainfall observations}
\label{Sec_2}
The rainfall data used in this study is based on satellite rainfall estimates of the Climate Prediction Center Morphing Method \citep[CMORPH,][]{Joyce2004}. The dataset has a spatial resolution of \ang{0.25} and covers the global area from \ang{60}S to \ang{60}N ({\textcolor{blue}{figure} \ref{totprecip_yearly}}). The temporal resolution is three hours, the time period used is 1998 to 2013.

The morphing method uses motion vectors which are derived from half-hourly interval geostationary satellite infrared imagery to propagate the precipitation estimates derived from passive microwave scans. In addition, the shape and intensity of the precipitation features are modified during the time between microwave sensor scans by performing a time-weighted linear interpolation. This process yields a spatially and temporally complete microwave-derived precipitation analysis, independent of the infrared temperature field. The dataset shows substantial improvements over both, simple averaging of the microwave estimates and over techniques that blend microwave and infrared information. Yet there are still several issues; one is the lack of accuracy of precipitation estimation over snowy regions. Another is related to precipitation that dissipates over regions that are not covered by the passive microwave satellite. This rainfall can not be captured by the morphing algorithm. Although CMORPH ranges among the best available satellite based rainfall estimates the correlation with rain gauge products is far from being perfect \citep{Joyce2004}.
\begin{figure}
	\centering
	\includegraphics[width=0.49\textwidth]{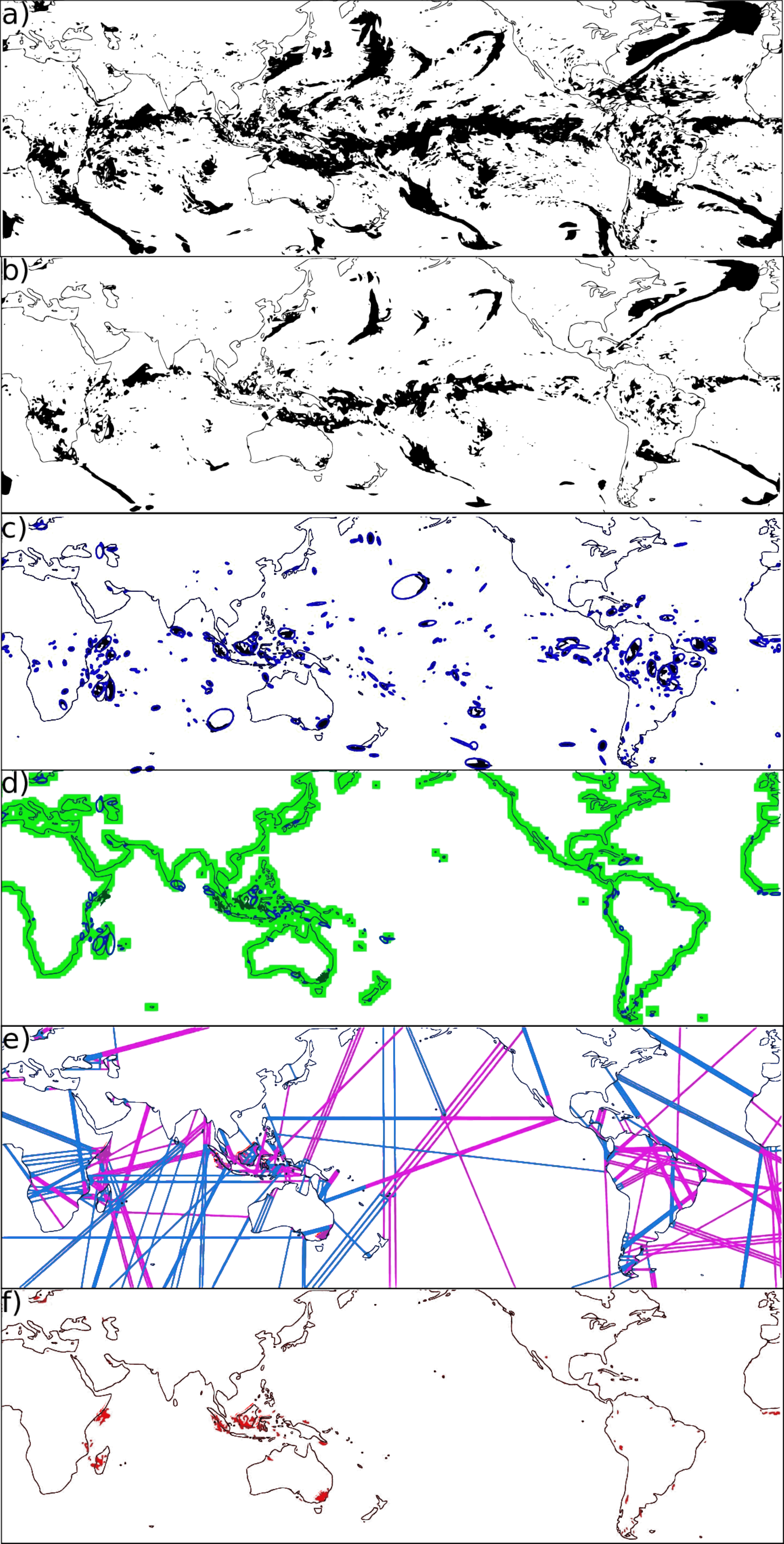}
\caption{The sequence of coastline induced rainfall recognition algorithm. a) Conversion of the rain data to a binary image. b) Applying a rainfall threshold. c) Deleting synoptic scale rainfall patterns and fitting the remaining patterns to ellipses. d) Defining a broad coastal area and deleting any rainfall domains not located in the green marked area. e) Alignment testing with straight lines for the fit-ellipse to the next coastline. f) Result.}
\label{Scheme}
\end{figure}
\section{Pattern Recognition}
\label{Sec_3}
In the present work we wish to extract rainfall whose structure indicates that it is associated with coastal phenomena like land-sea breezes convergence. Previous studies have shown that rainfall related to coastal land-sea interaction is mainly of higher intensity than the rainfall that isn't affected by the presence of the coastline in that area. Furthermore the precipitation patterns of interest are aligned with the coastline and should occur in the vicinity of the coastline \citep[e.g.,][]{Mori2004,Keenan2008,Qian2008,Hill2010}. Based on this knowledge we define four heuristics that have to be met to identify precipitation as associated with coastal land-sea interaction:

\begin{enumerate}
\item compared with the surrounding precipitation, the rainfall of interest has a higher intensity,
\item the recognized coastal precipitation is not large-scale,
\item rainfall due to land-sea interaction occurs within $\approx$ \SI{250}{\kilo\meter} of the coast, 
\item the precipitation pattern is aligned with the coastline. 
\end{enumerate}
The above heuristics are applied in 6 steps which we briefly summarize below and in \textcolor{blue}{figure} \ref{Scheme}a - \ref{Scheme}f. A more technical description of our algorithm can be found in the \textcolor{blue}{appendix}.

\begin{enumerate}
\item Application of a rainfall threshold to the rain data.

To apply the first mentioned heuristic, we choose local monthly rainfall percentile thresholds instead of a hard-threshold. At every grid-point and for every month we determine the deciles of the 3-hourly rainfall distribution and apply thresholds of the 20th, 50th and 80th percentile respectively (see below). With monthly percentiles as a threshold regional and seasonal variations of the rainfall are taken into account.

\item Converting the rainfall data to binary data.

Here any rainfall greater than the rainfall threshold is set to 1, lower values are assigned as 0. To test geometrical aspects like size and coastline alignment each contiguous rainfall area has to be considered separately. The separation separation of the different rainfall areas uses  a gradient method, which performs best on binary data (see \textcolor{blue}{appendix} for further details).

\item Deletion of large-scale rainfall.

After separation the total size of each contiguous rainfall area is measured. Using a size threshold of \mbox{$10^6$\SI{}{\kilo\meter}} rainfall patches that can be accounted as synoptic scale patterns, like large frontal systems, are then deleted from the data.

\item Fitting the rainfall domains with ellipses and applying an eccentricity threshold.

Since we are interested in rainfall features that stretch along a coastline we assume that the detected features are elongated. Therefore the rainfall domains are least-square fitted with ellipses. After fitting each contiguous rain area is enclosed by an ellipse. We then apply a threshold for the eccentricity of the ellipses to delete rainfall patterns that aren't elongated. The application of an eccentricity threshold is also necessary because the remaining rainfall patterns will be tested for alignment with the coastline. Alignment can only sensibly be determined if the rainfall area has a clear orientation. This is ensured by an eccentricity of the fit-ellipse larger than a threshold.

\item Deleting rainfall not occurring in the vicinity of the coast.

Here we define a broad area (\mbox{$\approx$ \SI{250}{\kilo\meter}}) around the coastline and delete all contiguous rainfall areas domains not occurring in the defined coastal area. Specifically, if less 20\% of the contiguous rainfall area lies within in the coastal vicinity it is deleted from the data set. With this blurred threshold we do not cut-off rainfall domains stretching over the borders of the coastal area. We are also able to detect patterns that are slightly further on- or offshore than the distance threshold defined in the heuristics. This has the advantage that the hard areal threshold becomes smoother. 

\item Testing the alignment of the rainfall domains with the coastline.

For alignment testing we first define a narrow coastal area (\mbox{$\approx$ \SI{50}{\kilo\meter}}). Any rainfall area with at least 90\%  of its area in this narrow coastal strip assumed to be aligned with the coastline and already marked as a coastally influenced precipitation patterns.  
\begin{table}
\caption{Threshold values used in the recognition algorithm yielding an ensemble of $3^3 = 27$ coastal rainfall estimates.}
\begin{tabular}{p{\mytable} r r r}\toprule
threshold for: & \multicolumn{3}{c}{values:} \\ \midrule
rainfall intensity [percentile] & 20 & 50& 80 \\ 
eccentricity [ ] & 0.2& 0.5& 0.8 \\
variation of straight line length[\%] & 5 & 25 & 50 \\ \bottomrule
\end{tabular}
\label{Threshold}
\end{table}
For each of the remaining objects three straight lines, two from the tips and one from the center of the fit-ellipse, are drawn orthogonally in both directions of the major axis. The distance from the origin to the next coastline intersection for each straight line is measured. If the standard deviation of the three distances is below a certain percentage of the mean distance, the contour is assumed to be aligned with a coastline. The object is then marked as aligned with the coast if the mean distance of all corresponding straight lines is not greater than \mbox{\SI{500}{\kilo\meter}}.
\end{enumerate}
All rainfall domains that have not been deleted during the 6 steps meet the previously mentioned heuristics and are assumed to be precipitation that is associated with coastal land-sea interaction. The algorithm is applied to all single three-hourly time steps of the CMORPH dataset from 1998 to 2013. A more comprehensive technical description of the algorithm can be found in the \textcolor{blue}{appendix}. The source-code of the algorithm can be downloaded freely via GitHub (\url{http://dx.doi.org/10.5281/zenodo.18173})

From the above discussion it is evident that thresholds have to be applied to run the pattern recognition algorithm. \begin{flushleft}
These thresholds are:\end{flushleft}
\begin{enumerate}
\item the area of the rainfall patches,
\item the intensity of the rainfall,
\item the eccentricity of the fitted ellipses,
\item the threshold for the standard deviation of different straight lines from the main axis of the fit-ellipse to the next coastline
\end{enumerate}
As the synoptic scale is well defined, the area threshold is set to a single value (\mbox{$\approx$ \mbox{$10^6$\SI{}{\square\kilo\meter}}}). The choice of the other three thresholds is more difficult.  We make reasonable choices for all three thresholds and generate an ensemble of 27 members by using all possible combinations of the threshold settings summarized in Table \ref{Threshold}.  The decision to use an ensemble was made to keep the method objective, rather than subjectively choosing a "perfect" parameter set. For each parameter we chose the highest and lowest reasonable threshold and one value in between. 

The 20th percentile is chosen as a reasonable low rainfall intensity threshold. Values less than the 20th percentile have would be contradictory to the a priori assumption that coastal rainfall is of relative high intensity. Secondly with low rainfall thresholds the rainfall data becomes more noisy. This leads to a poor performance of the applied Canny-edge-detection.
\begin{figure*}
 \centering
	\includegraphics[width=\textwidth]{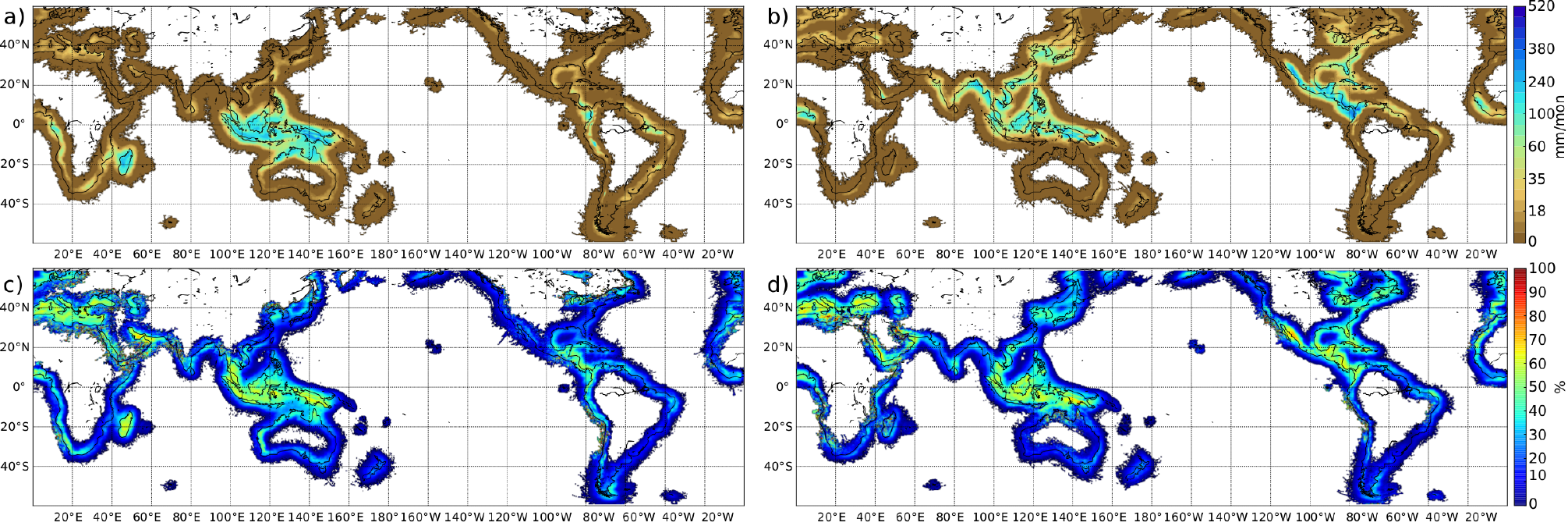}
	\caption{Mean coastal rainfall for DJF(a,c) and JJA (b,d). The two top panels show the ensemble mean of detected coastal rainfall, the two bottom panels show the detected precipitation as a percentage of total rainfall.} 
	\label{Overall} 
\end{figure*}
Canny-edge detection performed best by applying a rainfall thresholds greater than the 80th percentile. With such high rainfall thresholds on the other hand only a small number of rainfall events are remaining. Therefore we chose to set the highest reasonable threshold choice as the 80th percentile. The application of the eccentricity threshold as an orientation parameter is similar. Low values indicate rainfall patterns that are rather round. To assign a clear major axis that is needed for alignment testing, the objects shouldn't be round but elliptical. We therefore chose 0.8 as a maximal threshold. Thresholds of less than 0.2 turned out to be effective for the alignment testing. On the other hand too much precipitation was deleted when further decreasing the threshold. Therefore we chose 0.2 as a reasonable lower threshold. 
Rainfall patterns with an angle of less than 45$^\circ$ can be considered as aligned with the coastline. The would roughly be corresponding to the variation of the straight lines by 50\%. Therefore we choose the upper 50 \% as an upper threshold. Nevertheless the testing with different straight length variation thresholds indicates that this parameter has less impact on the results than the threshold choice for rainfall intensity and eccentricity of the fit-ellipse.

\section{Overall climatology of detected coastal rainfall}
\label{Sec_4}
All results presented here are based on the ensemble of objectively identified coastal rainfall described in the previous section.  All ensemble members are taken into account and no weighting is applied.

The climatology of the detected coastal precipitation, shown in \textcolor{blue}{figure} \ref{Overall}\textcolor{blue}{a} and \ref{Overall}\textcolor{blue}{b}, clearly reveals the expected seasonal variability in the tropics and subtropics. During DJF coastal rainfall maxima occur in the Maritime Continent region and over Madagascar. JJA shows peaks over the Bay of Bengal, the central American Pacific coast, the North American Atlantic coast and tropical West Africa. Also relatively high amounts of precipitation are detected over the Gulf of Carpenteria (\mbox{\ang{15}S,\ang{140}E}) during austral summer. In both seasons high amounts of coastal rainfall are detected in the Bight of Panama and around the Maritime Continent. 
  
In the tropical coastal regions high values of coastal rainfall are accompanied by a high percentage of the contribution of coastal rainfall to the total (\textcolor{blue}{figure} \ref{Overall}\textcolor{blue}{c} and {\ref{Overall}\textcolor{blue}{d}). Perhaps the most prominent example all year round is the Maritime Continent, with the coastal rainfall contribution frequently exceeding 50\% and approaching 66\% in some regions. This highlights the importance of land sea interaction for Maritime Continent rainfall. Relatively high amounts of coastally related rainfall are found in the highlands of New Guinea. This might originate from orographic effects caused by the mountain flanks being roughly parallel to New Guinean coastline. Interaction of coastal and orographic effects are difficult to separate. \cite{Qian2011} showed the strong interconnection of orographic and coastal effects for the Maritime Continent rainfall. Furthermore, \cite{Mapes2003b} conjectured a complex interaction of stratified air above topography and moist convection triggered by sea-breeze lifting. Therefore we account these detected features as coastline associated.
\begin{figure*}
\includegraphics[width=0.95\textwidth]{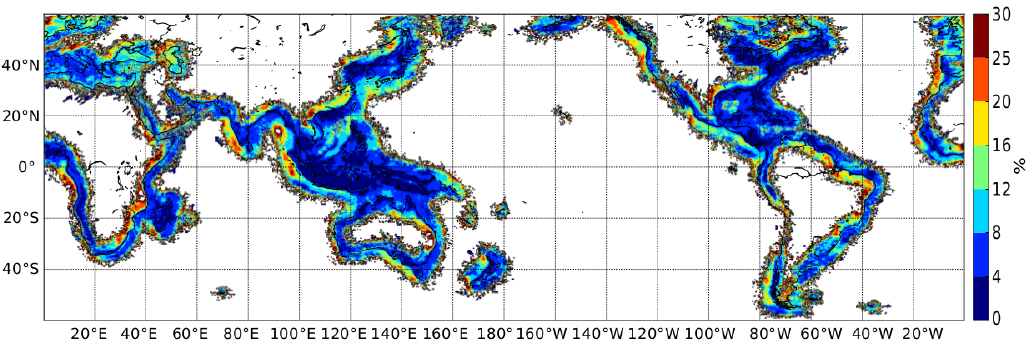}
\caption{The ensemble standard deviation as  percentage of the ensemble mean}
\label{Ensemble_std}
\end{figure*}
The fraction of detected coastal rainfall also reveals the importance of land-sea interaction in arid- and semi-arid coastal regions. The Red Sea, the Persian Gulf, the Australian west coast and the south west African coast are examples where despite relatively modest overall rainfall amounts, major portions of the precipitation are aligned with the coast. The Mediterranean is the second largest region (after the Maritime Continent) that is strongly influenced by land-sea interaction. Even in the northern hemisphere winter the percentage of detected coastal precipitation remains high. This signal during DJF in the Mediterranean might have its origin in frontal systems that are aligned with the coastline and should therefore be treated with caution.

{\textcolor{blue}{figure} \ref{Ensemble_std}} shows the ensemble standard deviation as a percentage of the ensemble mean. It is interesting to note that the ensemble members show better agreement over tropical land areas than over the adjacent ocean. The standard deviation tends to rise with an increasing distance from the coastline. It is known that with greater distance from the coast the precipitation patterns become more influenced by ambient flow patterns \citep{Gilliam2004,Azorin-Molina2009}, thus the alignment of the rainfall areas with the coast becomes more sensitive to the application of an alignment threshold. Relatively high standard deviations are found in the extra tropics. A more detailed investigation reveals that in this area our algorithm is particularly sensitive to the application of the rainfall threshold, leading to larger variations across the different ensemble members.

\section{The diurnal cycle of coastal rainfall}
\label{Sec_5}
 \begin{figure*}
\includegraphics[width=\textwidth]{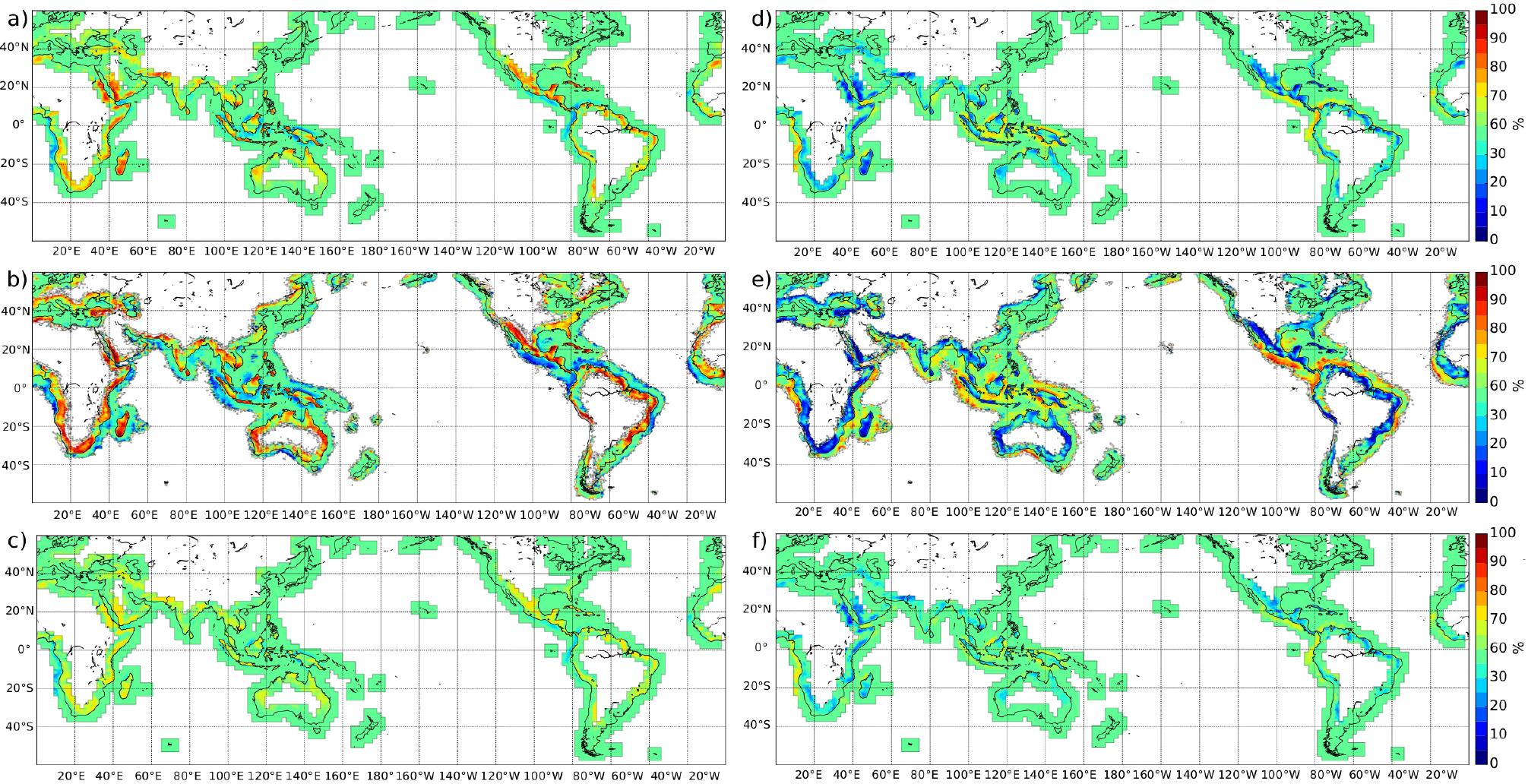}
\caption{Mean annual (left) daytime and (right) nighttime rainfall as a fraction of overall daily rainfall for (top) total precipitation, (middle) detected coastal precipitation, and (bottom) residual rainfall. Daytime is defined as 0900-2100 LT and nighttime as 2100-0900 LT. For (top) and (bottom) only, rainfall within 250 km onshore and offshore is shown.}
\label{DNPerc}
\end{figure*} 

\textcolor{blue}{figure} \ref{DNPerc} shows the local daytime (left) and nighttime (right) rainfall as a fraction of overall rainfall for all rainfall (top), the detected coastal rainfall events only (middle) and the residual rainfall considered non-coastal in nature (bottom). Here, daytime is referred to as 0900 - 2100 LT and nighttime as 2100 - 0900 LT. For total rainfall (\textcolor{blue}{figure} \ref{DNPerc}\textcolor{blue}{a}, {\ref{DNPerc}\textcolor{blue}{d}) the diurnal variability is very large in the Maritime Continent, the Bight of Panama, the west coast of Central America, the eastern Bay of Bengal, the African East Coast and around the Horn of Africa. The fraction of detected daytime rainfall, shown in \textcolor{blue}{figure} \ref{DNPerc}b} and {\ref{DNPerc}e}, reveals a more pronounced diurnal variation than the total rainfall. Some regions have up to 95\% more rainfall during the day over land than over the adjacent sea and vice versa for nighttime rainfall. The most significant diurnal rainfall variations occur over the Maritime Continent, Madagascar and the Bight of Panama. The fraction of local nighttime rainfall to overall rainfall is shown on the right hand side of \textcolor{blue}{figure} \ref{DNPerc}}. Note that day- and nighttime rainfall together add to total rainfall. Strong nighttime signatures of detected coastal precipitation are evident over the central American Pacific coast and the east coast of the United States as well as the Maritime Continent. Weaker but still discernible nightitme signals are detected in south east Asia, the Bay of Bengal and around the Philippines. Comparing the daytime and nighttime fractions it is evident that the diurnal rainfall cycle is stronger over land than over the adjacent ocean.

It is well known that the coastal diurnal rainfall cycle is strongly affected by land-sea interaction. Therefore we hypothesize that a) diurnal rainfall variations of the detected coastal precipitation should be much larger than those of the overall rainfall near coasts and b) the residual rainfall that hasn't been detected as coastally influenced should reveal very little diurnal variation. \textcolor{blue}{figure} \ref{DNPerc}\textcolor{blue}{c} shows this remaining residual day- and nighttime rainfall as a fraction of overall residual rainfall. The residual fractions of daytime and nightime precipitation are small with values mostly between 35 - 65 \%. Note that no diurnal variation would be indicated by 50\% (green areas). This supports our claim that the presented technique is able to capture the majority of rainfall that is due to land-sea interaction in coastal areas. In some areas the residual rainfall variation remains significant though. Especially the north west coast of Borneo, the west coast of Sumatra and the Bight of Panama show a relatively high amount of residual nighttime rainfall over the ocean.

\section{Coastal rainfall over the Maritime Continent}
\label{Sec_6}
\begin{figure}[t]
\centering
\includegraphics[width=0.5\textwidth]{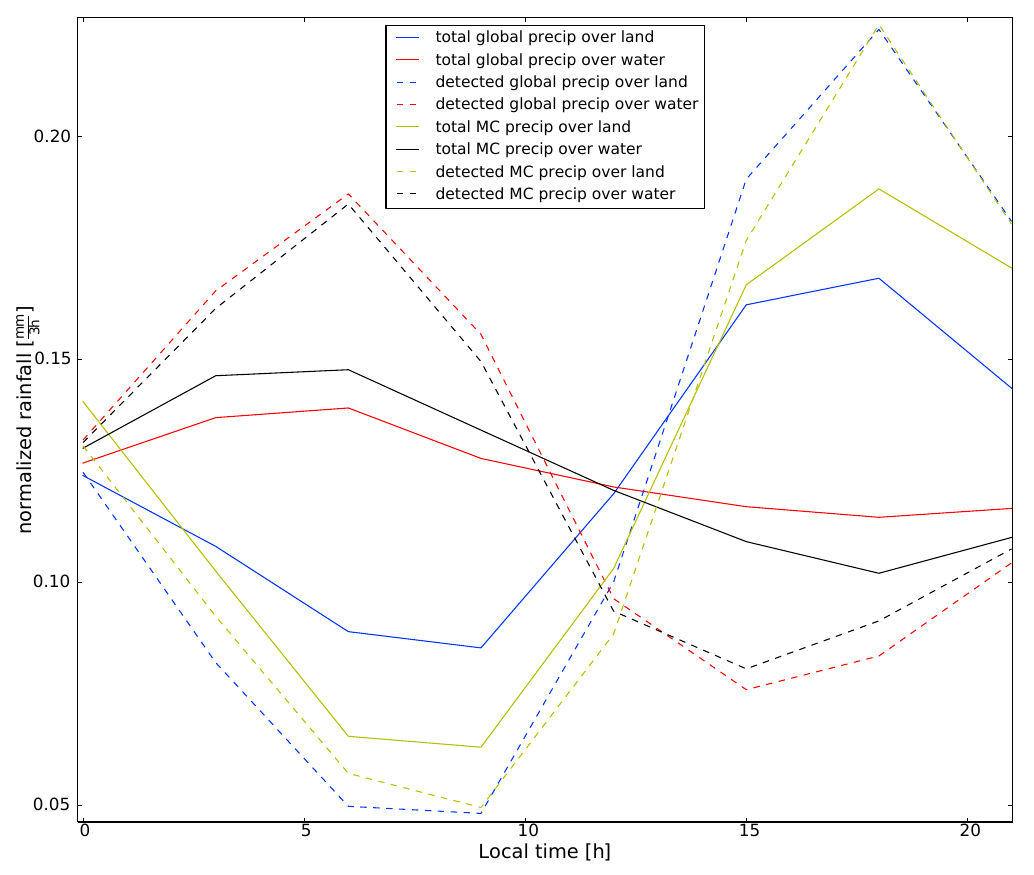}
\caption{The mean occurrence of global rainfall and rainfall over the Maritime  Continent over land and ocean as a function of the time of the day. For a better comparison of the diurnal global and Maritime Continent rainfall the sum of a full rainfall cycle is set to one.}
\label{Timing} 
\end{figure}

The Maritime Continent has, not surprisingly, been identified as one of the regions where precipitation is particularly strongly influenced by land-sea interaction. 
\textcolor{blue}{figure} \ref{Timing} shows the average timing of rainfall over land and over water for the Maritime Continent and on global average. Here, the dashed lines represent the detected coastal precipitation and the solid lines the total precipitation. For a better comparison the rainfall is normalized by making the sum of a full cycle equal to one.  The diurnal cycle over the Maritime continent is clearly more pronounced than on global average. This is especially true for the land regions. The overall distribution of the maxima shows the expected behavior. Over land the rainfall peaks in the late afternoon whereas the maximum over water occurs in the early morning. This is in agreement with the results of \cite{Mori2004} who were using TRMM-3B42 satellite rainfall estimates.

The detected precipitation shows a stronger diurnal variation, especially over water. The detected minimum of oceanic rainfall occurs roughly three hours earlier than the total rainfall minimum. In general the detected coastal precipitation variation is most pronounced over land. 
The detected coastal rainfall over the Maritime Continent shows roughly the same diurnal variation as the recognized precipitation on global average.

\begin{figure}
\includegraphics[width=0.5\textwidth]{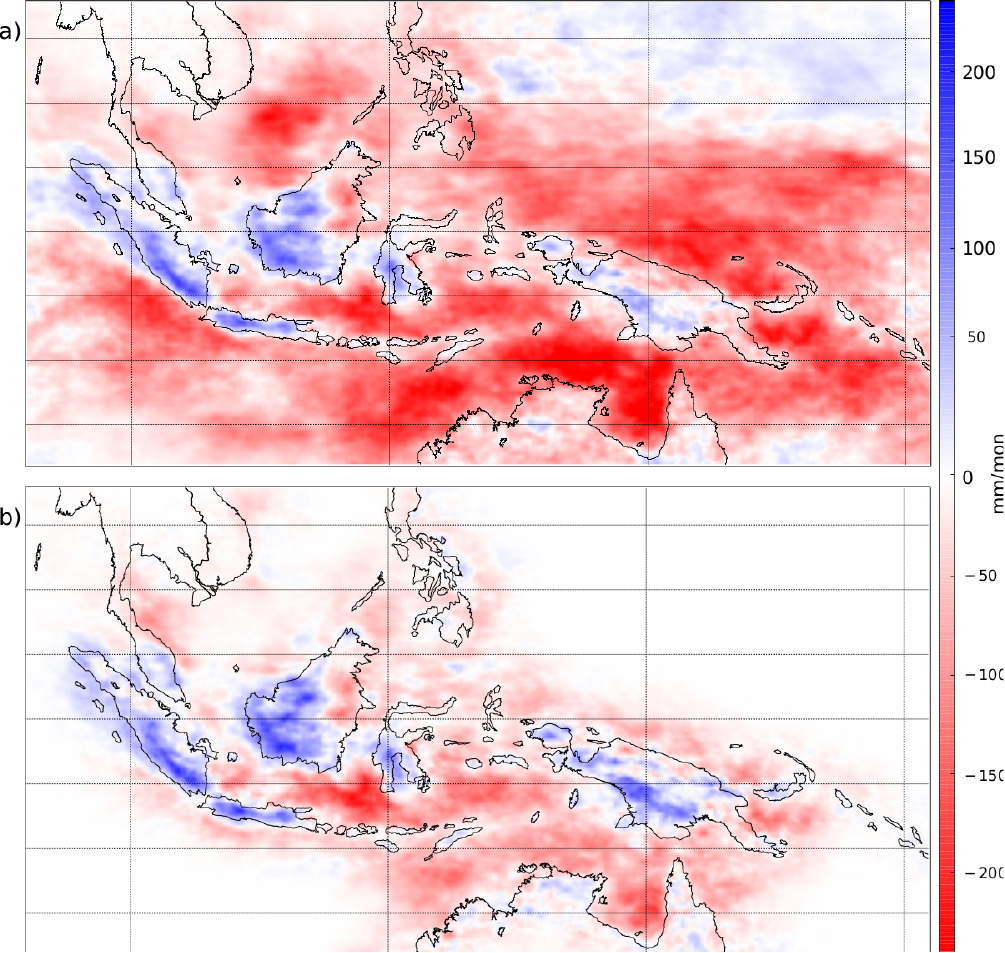}
\caption{Mean difference of DJF rainfall during suppressed and active MJO phase for a) total rainfall and b) detected coastal rainfall.}
\label{MJO}
\end{figure}

To further assess the utility of the coastal rainfall dataset, we investigate the influence of large scale modes of climate variability on coastal precipitation in the Maritime Continent. In particular, we follow on from earlier studies \citep{Rauniyar2010,Rauniyar2012} and investigate how the MJO and ENSO modulate the diurnal cycle. First, the difference of coastal rainfall in the Maritime Continent region during active and suppressed MJO phases is calculated. The MJO is considered to be active over the Maritime Continent in phases 4-6 of the Wheeler-Hendon real-time multivariate MJO (RMM) index \citep{Wheeler2004}. The inactive phase is defined with index values of 1,2,7,8. Only days with an RMM amplitude of equal or greater 1 were sampled. \textcolor{blue}{Figure} \ref{MJO} shows the difference of total (a) DJF rainfall and DJF detected (b) coastal rainfall between suppressed and active MJO phase.

\begin{figure}
\includegraphics[width=0.5\textwidth]{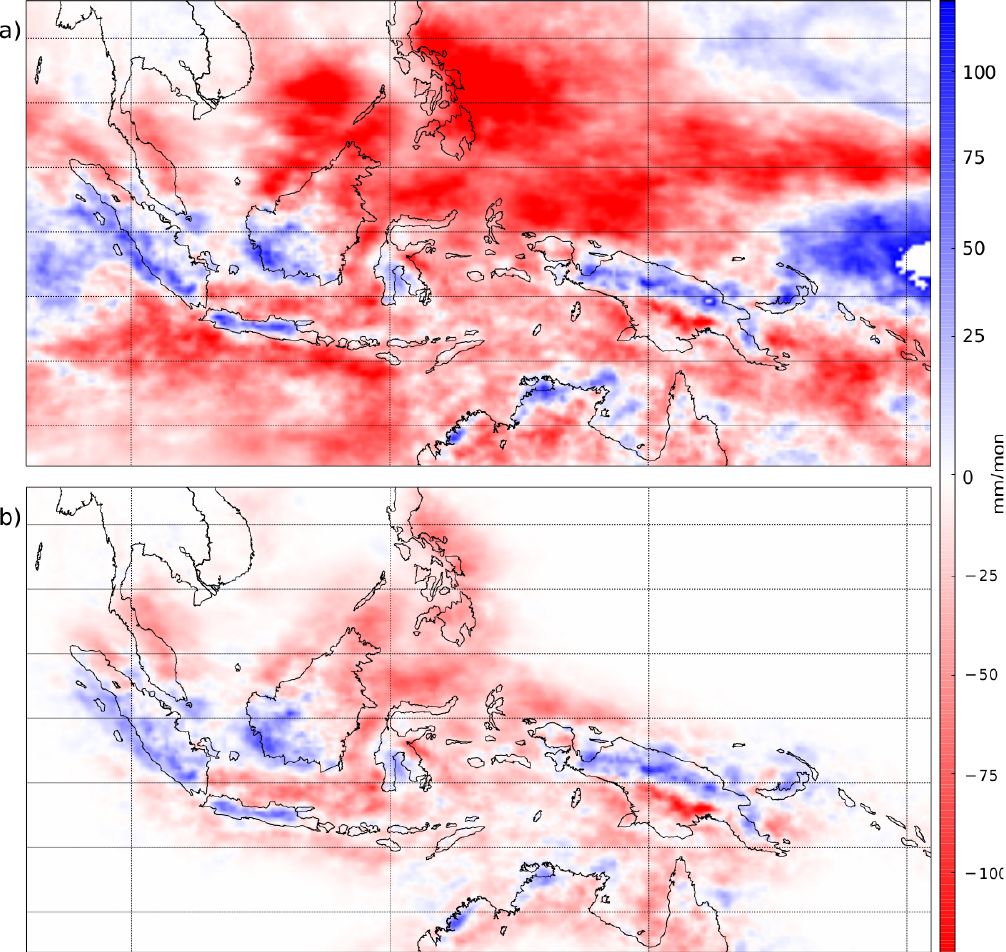}
\caption{mean difference of DJF rainfall during weak ENSO and El Ni\~{n}o phase for a) total rainfall and b) detected coastal rainfall.}
\label{ENSO}
\end{figure}

In general there is more rainfall during the active phase over the ocean. However, as has been found in previous
studies like \cite{Rauniyar2010} who where using \textit{non}-objective methods, over land, there are large areas where rainfall during the suppressed phases is larger than that during the active phase. Interestingly both the magnitude and pattern of the overall signal are well reproduced when considering only coastally influence precipitation. This leads us to conclude that it is mainly the precipitation due to land-sea interaction that leads to an increase of rainfall to over land during the suppressed phase. This is an important finding especially in the light that weather and climate models of coarse resolution are unlikely to be able to capture land-sea effects correctly.
The El Ni\~{n}o Southern Oscillation (ENSO) also influences the rainfall over the Maritime Continent. Global rainfall pattern differences tend to dryer conditions in the western Pacific during the El Ni\~{n}o phase of the oscillation. ]\textcolor{blue}{Figure} \ref{ENSO}\textcolor{blue}{a} shows the mean DJF difference of total rainfall between El Ni\~{n}o years and weak ENSO years. Our El Ni\~{n}o definition is based on the Oceanic Ni\~{n}o Index (ONI). The ONI is based on a 3 monthly running mean of the SST anomaly in the Nino3.4 region (\ang{5}N-\ang{5}S, \ang{120}E-\ang{170}W). ONI values of greater than 0.5 indicating El Ni\~{n}o events and values within {$\pm$ 0.5} showing weak ENSO episodes.

It can be clearly seen that there is more rainfall over the ocean during years in which ENSO is weak. On the islands on the other hand the situation looks different. There is more rainfall over land regions of the Maritime Continent during El Ni\~{n}o years. The rainfall pattern difference for the detected coastal rainfall looks very similar to the total precipitation (see \textcolor{blue}{figure} \ref{ENSO}). The magnitudes of the positive detected coastal rainfall anomalies over land are about the same order as the anomalies of the total rainfall. 

Once again, as for the MJO events, the variations of rainfall with El Ni\~{n}o support the hypothesis that during suppressed convective large scale conditions, coastal effects strongly modulate the rainfall over the Maritime Continent and contribute largely to an enhancement of rainfall \citep{Qian2013,Rauniyar2012}. The fact that the results using the \textit{objectively} detected coastal rainfall over the islands are very similar to those using total rainfall provides some evidence that the algorithm developed in this study is able to identify coastally driven precipitation events well.
\section{Summary and Conclusion}
\label{Sec_7}

In the present study a pattern recognition technique is developed and applied to a global three-hourly rainfall dataset to detect rainfall associated with land-sea interactions. The technique applies thresholds for four different characteristics to extract rainfall that is likely driven from coastal features. Since the algorithm employs several thresholds and since the implications of the threshold choice are not entirely knowable an ensemble of 27 different threshold setups is created.

The ensemble mean of the generated dataset reveals the expected seasonal and diurnal variability of coastal tropical precipitation. The standard deviation among the ensemble members, which is on average 4\% of the ensemble mean, indicates the overall robustness of the algorithm. Over the Maritime Continent and the Bight of Panama major portions of total precipitation can be clearly related to coastal land-sea interaction. High fractions of rainfall associated with coastal rainfall over the Mediterranean, the Red Sea, Persian Gulf, the South African and Australian west coast highlight the importance of coastal processes in arid and semi-arid regions of the world. Most of the expected features of coastal convection are captured by the algorithm. This is indicated by the strong diurnal cycle of the detected rainfall and relatively weak diurnal rainfall variation of the residual rainfall.

The coastal rainfall recognition method is only based on a few geometrical aspects of precipitation, no further criteria have to be met. This differs from many statistical approaches, like cluster, spectral or principal component analysis where assumptions like the occurrence of diurnal harmonics have to be applied. Therefore already known statistical properties, like the diurnal precipitation cycle, can be studied to evaluate the newly developed method. Moreover further aspects of coastal precipitation can now be easily illustrated. For instance coastline effects are known to be an important trigger for deep precipitating convection \citep{Simpson1980,Simpson1993,Qian2008}. The presented method reveals that up to two thirds of the total Maritime Continent rainfall can be related to coastal effects. 

We are also able to \textit{objectively} show that coastline associated precipitation is influenced by modes of large scale variability like the MJO and ENSO. The results suggested that during suppressed large scale convective conditions the total precipitation over land areas is strongly modulated by coastline effects. This is in accordance with previous studies using \textit{non}-objective statistical methods \citep{Rauniyar2010,Rauniyar2012}. 

The strength of the method presented here is that it is objective and relatively straightforward to apply. This has allowed us, to our knowledge for the first time, to characterize coastal precipitation globally rather than just for a limited amount of regional cases. The dataset of coastal convection features will also be very useful for future studies. For instance, it can be applied to identify the background atmospheric state in which coastal precipitation occurs. This is crucial for the parametrization of such rainfall in models, as the MJO and ENSO analysis shows, that coastal effects likely allow rainfall to exist in large scale conditions, in which rainfall is suppressed over the open ocean. Since it can be driven with any kind of grided rainfall data the algorithm can moreover be utilized to evaluate coastal precipitation projected within climate models.

\begin{acknowledgment} 
We acknowledge the Australian Research Council's Centre of Excellence for Climate System Science (CE110001028) for funding this work. We would also like to thank Rit Carbone from NCAR and Chris Holloway from University of Reading for their useful suggestions and comments in the early stage of this work. Furthermore we thank the three anonymous reviewers for their valuable comments and suggestions to improve the quality of the publication. The CMORPH satellite based rainfall estimates were obtained from the Climate Prediction Center (CPC) of the National Oceanic and Atmosphere Administration (NOAA). The tools utilized for the pattern recognition are supplied by the open source image processing library OpenCV. The source code and a documentation can be retrieved from GitHub\\ (\url{http://dx.doi.org/10.5281/zenodo.18173})
\end{acknowledgment}
\ifthenelse{\boolean{dc}}
{}
{\clearpage}

\begin{appendix}
\section*{\begin{center}The pattern recognition algorithm\end{center}}
\label{appendix}
A key task in the mining of remote sensing imagery is the identification of static structures such as buildings, roads and bridges. \cite{Nevatia1982} were one of the first who identified the airports in the San Francisco Bay area. They applied some simple characteristics about the airports in this area. Specifically they assumed an airport has a straight runway of certain length and is located near the coast. Starting from this heuristics they were using simple image segmentation techniques to detect the target features. To detect the target precipitation features we also use image segmentation. Image segmentation is a technique where information is extracted from the background of an image until all a priori assumptions are met (see \textcolor{blue}{section} \ref{Sec_3} for the heuristics used here) and the remaining part of the image can be considered as the detected pattern.

We start the pattern detection by applying the rainfall intensity threshold and converting the rainfall data to a binary image. Areas less than the threshold are set to 0 and the ones above to 1. After converting the rainfall data to an binary image small holes within the rainfall areas (1-domains) are closed. Closing of small holes within a connected domain can be done by dilation and erosion. Dilation and erosion are methods where a domain of an image ($A$) is probed with a structuring element ($B$) and it is quantified how the element fits inside an image object. The structuring element, in image processing also referred as kernel, is applied on the binary image. In the present case a cross of \mbox{$3\times3$ pixels} is chosen as kernel shape. Formally dilation and erosion are defined as:

\begin{equation*}
\begin{split}
\mathrm{dilation: }\  & A \oplus B = \bigcup\limits_{b\in B} A_b \\
\mathrm{erosion: }\ & A \ominus B = \bigcap\limits_{b\in B} A_{-b}
\end{split}
\end{equation*}
For dilation the kernel $B$ is scanned over the image domain $A$, the pixels overlapped by $B$ are added to $A$. Thus, the black regions within $A$ are growing. Erosion is very similar to dilation, the center of $B$ is subtracted from $A$ if $A$ and $B$ are only partly overlapping. Therefore the black regions in $A$ are shrinking. These two methods are applied consecutively to close small holes within the domain $A$.

To apply the heuristics mentioned in \textcolor{blue}{section} \ref{Sec_3} each rainfall domain has to be considered separately. The separation of the patterns is realized by Canny-edge detection \citep{Canny1986}. For Canny-edge detection an image is Gaussian filtered to remove noise from the image. After de-noising the image local maxima of gradients within the domains that are to be separated and their surroundings are found. The Gaussian filter is defined as:

\begin{equation*}
G(i,j) = \dfrac{1}{2 \pi \sigma^2} \cdot \mathlarger{\mathlarger{e}}^{-\dfrac{(i-k)^2+(j-k)^2}{2\sigma^2}}
\end{equation*}
With:
\begin{itemize}
\item $\sigma$ the standard deviation of the Gaussian
\item $k$ the size of the kernel.
\end{itemize}
A standard deviation of 3 pixels combined with a kernel size of 3 pixel performed best for the application of the Gaussian filter. Eventually Canny-edge detection is applied on the resulting image. 

From now on every closed contour that has been detected is considered as an independent object. The first step after contour separation is to delete large domains. The remaining objects are fitted to ellipses. We chose ellipses as fit objects because geometric properties like orientation and aspect ratio of the fitted pattern are easily retrieved by the location of the main axis and the value of numerical eccentricity of the ellipse. The fit is based on least square fitting and assumes that all pixels within an object belong to one ellipse. Further details are provided in \cite{Fitzgibbon1999} and in \cite{Mulchrone2004}.

The next step includes the a priori assumption that the target patterns occur no further than roughly \SI{250}{\kilo\meter} on- or offshore. For this purpose we define a a coastline by applying Canny-edge detection to a land-sea mask of the same resolution as the data. The coastal area is defined by adding pixel to the neighbourhood of the coastal edges. This technique can be seen as the reversal of box counting that is applied to define fractal geometries such as coastlines \citep{Block1990}. Finally all domains that have less than 80\% of their total area within the defined area are deleted from the image.

The remaining patterns are tested for alignment with the coastline. Alignment can only be tested if each fit-ellipse has an assignable major and minor axis. This is guaranteed if the numerical eccentricity of the fit-ellipse is considerably larger than 0. Therefore all domains with an eccentricity of the fit-ellipse lower than a certain threshold are deleted.
The alignment with the coastlines of the remaining domains is tested in three ways. First a narrow coastal area ($\approx 2 \times $resolution), is defined by inverse box counting. Domains having more than 90\% of their area overlapping with the narrow coastal area are considered as aligned with the coastline. The remaining domains are shifted into the narrow coastal area without changing their orientation. All domains that are again overlapping by at least 90\% are marked as detected. For each of the remaining objects three straight lines, two from the tips and one from the center of the fit-ellipse, are drawn orthogonally in both directions of the major axis. The distance from the origin to the next coastline intersection for each straight line is measured. If the standard deviation of the three distances is below a certain percentage of the mean distance, the contour is assumed to be aligned with a coastline. The object is then marked as aligned with the coast if the mean distance of all corresponding straight lines is not greater than \SI{500}{\kilo\meter}. 

All domains that haven't been deleted meet all in \textcolor{blue}{section} \ref{Sec_3} mentioned heuristics and are finally labelled as coastally influenced rainfall features.







\end{appendix}

\ifthenelse{\boolean{dc}}
{}
{\clearpage}
\bibliographystyle{ametsoc2014}
\bibliography{references}

\end{document}